\documentclass[10pt,superscriptaddress,twocolumn,amsmath,amssymb,aps,pra,showpacs]{revtex4-1}
\usepackage{mathrsfs}
\usepackage{graphicx}
\usepackage{dcolumn}
\usepackage{bm}
\usepackage[title,titletoc,header]{appendix}

\usepackage{amssymb}
\usepackage{amsmath}
\usepackage{mathrsfs}
\usepackage{amsmath}
\usepackage{subfigure}
\usepackage{float}
\usepackage[title,titletoc,header]{appendix}
\usepackage{graphicx}

\newcommand{\be}{\begin{equation}}
\newcommand{\ee}{\end{equation}}
\newcommand{\bea}{\begin{eqnarray}}
\newcommand{\eea}{\end{eqnarray}}

\begin{document}
\title{Lyapunov exponent, mobility edges  and critical region in the generalized Aubry-Andr\'{e}  model with an unbounded quasi-periodic potential}
\author{Yi-Cai Zhang}\thanks{Corresponding author. E-mail:~
zhangyicai123456@163.com}
\affiliation{School of Physics and Materials Science, Guangzhou University, Guangzhou 510006, China}

\author{Yan-Yang Zhang}
\affiliation{School of Physics and Materials Science, Guangzhou University, Guangzhou 510006, China}


\date{\today}
\begin{abstract}
In this work, we  investigate  the Anderson localization problems of the generalized Aubry-Andr\'{e}  model (Ganeshan-Pixley-Das Sarma's model) with an  unbounded quasi-periodic potential where the parameter $|\alpha|\geq1$. The Lyapunov exponent $\gamma(E)$ and the mobility edges $E_c$ are exactly obtained for the unbounded quasi-periodic potential.
With the Lyapunov exponent, we find that there exists a critical region in the parameter $\lambda-E$  plane.  The critical region consists of critical states.
 In comparison with localized and extended states, the fluctuation of spatial extensions of the critical states is much larger. The numerical results show that the scaling exponent of inverse participation ratio (IPR) of critical states $x\simeq0.5$.
Furthermore,  it is found that the critical indices of localized length  $\nu=1$ for bounded ($|\alpha|<1$) case and $\nu=1/2$ for unbounded ($|\alpha|\geq1$) case.
The above distinct critical indices can be used to distinguish the localized-extended from localized-critical transitions.
At the end, we show that the systems with  different $E$ for both cases of $|\alpha|<1$ and $|\alpha|\geq1$
can be classified by the Lyapunov exponent $\gamma(E)$ and Avila's quantized acceleration $\omega(E)$.

\end{abstract}

\maketitle
\section{Introduction}
For conventional orthogonal class system, it is believed that an arbitrarily weakly uncorrelated diagonal disorder in one and two dimension \cite{Economou} can result in the Anderson localization \cite{Anderson1957}, that all the eigenstates are localized. However, in the presence of off-diagonal disorders, one-dimensional system can have extended states \cite{Theodorou1976,Antoaiou1977}.
In three dimension, there exist mobility edge $E_c$ which separates the localized from extended states \cite{Economou1972}. When the energies approach the mobility edge $E_c$, the localized length of localized states  would diverge.

In one dimension, a famous example where the localized-extended transition can occur is the Aubry-Andr\'{e} lattice  model (AA model) \cite{Aubry1980}, i.e.,
\begin{align}\label{10}
t[\psi(i+1)+\psi(i-1)]+2\lambda \cos(2\pi\beta i+\phi)\psi(i)=E\psi(i).
\end{align}
where $t$ is hopping, $i\in Z$ is lattice site index, $2\lambda$ describes the quasi-periodic potential strength, $\beta$ is an irrational number, $\phi$ is a phase.
When the quasi-periodic potential is weak, i.e., $|\lambda/t|<1$, all the eigenstates are extended states. When the potential strength is sufficiently large  ($|\lambda/t|>1$), all the eigenstates become localized states with a localized length $\xi=1/\ln(\lambda/t)$.
At critical point ($|\lambda/t|=1$), all the eigenstates are critical states. So there is no mobility edges in the AA model. The non-existence of mobility edges originates from exact self-duality of this model at the critical point.
In general, the breaking of the self-duality would result in the appearance of mobility edges in the one dimension system \cite{Delyon1984,Sarma1988,Sarma1990,Izrailev1999,Boers2007, Sil2008,Liu2017,Li2017,Luschen2018,Tang2021}.

A generalized Aubry-Andr\'{e}  model (GAA model) which can have mobility edges has been proposed by Ganeshan, Pixley and Das Sarma  \cite{Biddle2010, Ganeshan2015,Duthie2021,Bodyfelt2014,Danieli2015}. The GAA model is
 \begin{align}\label{10}
t[\psi(i+1)+\psi(i-1)]+\frac{2\lambda \cos(2\pi\beta i+\phi)}{1-\alpha \cos(2\pi\beta i+\phi)}\psi(i)=E\psi(i).
\end{align}
In comparison with the AA model, there is an extra parameter $\alpha$ which is a real number.
Surprisingly, the mobility edges can be exactly obtained with a generalized self-dual transformation. Later, the mobility edges have been experimentally observed \cite{An2021}.
Very recently, a so-called mosaic model has been proposed  \cite{Wangyucheng2020} which also has mobility edges and localized-extended transitions. The Lyapunov exponent and mobility edges can be exactly obtained with Avila's global theory on the single frequency quasi-periodic potentials \cite{Liu2021,Avila2015}.

In the previous studies (for example in Refs. \cite{Ganeshan2015,Liu2021}), the parameter $\alpha$ in GAA model is limited to $|\alpha|<1$ due to the concerns of the possible appearance of divergences in the quasi-periodic potential [see Eq.(2)].
A natural question arises, aside from the unboundedness of potential, how about is it if $|\alpha|\geq1$?
 One may wonder whether there exist mobility edges for  $|\alpha|\geq1$. What are the localized properties of eigenstates?

In this work, we try to answer the above questions by extending the previous investigations of GAA model into a regime where parameter $|\alpha|\geq1$. It is found that there are also mobility edges $E_c$. The Lyapunov exponent $\gamma(E)$ and mobility edges are also exactly obtained with the Avila's theory.
In addition, we find that when $|\alpha|\geq1$, in the parameter $(\lambda, E)$  plane, a critical region which consists of critical states would appear. In comparison with the localized and extended states, the extensions of eigenstates in the critical region have much larger fluctuations.
Near the mobility edges, there exist localized-critical transitions where the localized length becomes infinite, e.g.,
\begin{align}\label{10}
\xi(E)\equiv1/\gamma(E)\propto|E-E_c|^{-\nu}\rightarrow\infty, \ \ as \ E\rightarrow E_c,
\end{align}
where the critical index \cite{Huckestein1990} $\nu=1/2$, which is different from that ($\nu=1$) of the case of $|\alpha|<1$.
Finally, we find that the systems with different parameter $E$ can be systematically classified  by Lyapunov exponent and Avila's acceleration.

The work is organized as follows. First of all, we discuss the properties of eigenenergies of Hamiltonian operator for both $|\alpha|<1$ and $|\alpha|\geq1$ in Sec.\textbf{II}.  In Sec.\textbf{III}, the Lyapunov exponents are obtained with Avila's theory.  Next, with the Lyapunov exponent, we  determine the mobility edges and critical region  in Sec.\textbf{IV}.
 At the end, a summary is given in Sec.\textbf{V}.

\
\begin{table}
\begin{center}
\begin{tabular}{|c|c|c|c|c|}
\hline
 lattice size &$N=100$& $N=300$ &  $N=500$&$N=700$ \tabularnewline
 \hline
 Max\{$|E_n|/t$\} &39.97& 64.44 & 402.97 &1186.65  \tabularnewline
 \hline
\end{tabular}
\end{center}
\caption{The unboundedness of energy spectrum for $\alpha=2$ and $\lambda=t$. We calculate the maximums of the absolute values of eigenenergies for lattice size $N=100,300,500$ and $N=700$, respectively. In our numerical calculations, we always take irrational number $\beta=\frac{\sqrt{5}-1}{2}$, phase $\phi=0$ and hopping $t=1$.
}
\end{table}

\section{bounded and unbounded energy spectrum of GAA model}
Eq.(2) can be viewed as an eigen-equation of Hamiltonian operator $H$, i.e.,
\begin{align}\label{10}
H|\psi\rangle=(H_0+V_p)|\psi\rangle =E|\psi\rangle,
\end{align}
where free particle part $H_0$ and potential $V_p$ in a second-quantized form  are
\begin{align}\label{10}
&H_0=t\sum_i[C^{\dag}_{i+1}C_{i}+C^{\dag}_{i}C_{i+1}],\notag\\
&V_p=\sum_i\frac{2\lambda \cos(2\pi\beta i+\phi)}{1-\alpha \cos(2\pi\beta i+\phi)}C^{\dag}_{i}C_{i},
\end{align}
where $C_{i}(C^{\dag}_{i})$ is the annihilation (creation) operator for state at site $i$.

\subsection{$|\alpha|<1$}
When $|\alpha|<1$, for an arbitrary integer  $i$,  due to $1-\alpha \cos(2\pi\beta i+\phi)>0$, the potential energy $\frac{2\lambda \cos(2\pi\beta i+\phi)}{1-\alpha \cos(2\pi\beta i+\phi)}$ is bounded.
So, the Hamiltonian $H$ is a bounded operator.  For an arbitrary state $|\psi\rangle$, the average value of energy $\langle H \rangle$ is finite, i.e., there exists a real number $M>0$, the relation
\begin{align}\label{10}
|\langle H \rangle|=\frac{|\langle\psi|H|\psi\rangle|}{\langle\psi|\psi\rangle}< M
\end{align}
holds. Consequently, all the eigenvalues $E_n$ of $H$ ($H|\psi_n\rangle=E_n|\psi_n\rangle$) are finite, i.e.,
\begin{align}\label{10}
|E_n|< M,
\end{align}
also holds.

\subsection{$|\alpha|\geq1$}
When $|\alpha|\geq1$,   due to the ergodicity of the map $\phi\longrightarrow 2\pi\beta i+\phi$  \cite{Arnold},  $|1-\alpha \cos(2\pi\beta i+\phi)|$  can be arbitrarily small if the lattice size is sufficiently large. Then the potential energy $\frac{2\lambda \cos(2\pi\beta i+\phi)}{1-\alpha \cos(2\pi\beta i+\phi)}$ can be arbitrarily large and
the Hamiltonian $H$ is an unbounded operator. So the average value of energy $\langle H \rangle$ is unbounded, i.e., for an arbitrary real number $M>0$, there exists a state $|\psi \rangle$, the relation
\begin{align}\label{10}
|\langle H \rangle|=\frac{|\langle\psi|H|\psi\rangle|}{\langle\psi|\psi\rangle}> M
\end{align}
holds. Consequently, the  set of eigenvalues $E_n$ of $H$ is also unbounded. Namely, for an arbitrary real number  $M>0$, there exists an eigenenergy  $E_n$, the relation
\begin{align}\label{10}
|E_n|> M
\end{align}
holds.

The above results have been verified by our numerical calculations. To be specific, we take total lattice site number $N>0$  and an $N\times N$ matrix associated with $H$ can be established with open boundary conditions at two end sites. Then we diagonalize it to get the $N$  eigenenergies and eigenstates.
In our whole manuscript, we use the units of $t=1$ and take irrational number $\beta=\frac{\sqrt{5}-1}{2}$ and phase $\phi=0$. For $\alpha=2$, we calculate the maximums of the absolute values of eigenenergies for lattice size $N=100,300,500$ and $N=700$, respectively. The results are reported in Table I.  From Table \textbf{I}, we see that the maximums of eigenenergies of $\alpha=2$  $(|\alpha|\geq1)$ grow rapidly with the increasing of lattice size $N$. It is  expected when lattice size $N\rightarrow \infty$, the range of eigenenergies would be infinitely large.

In addition, when $|\alpha|\geq1$ and the potential energy $\frac{2\lambda \cos(2\pi\beta i+\phi)}{1-\alpha \cos(2\pi\beta i+\phi)}$ is sufficiently large, the free particle part $H_0$ is negligible in Eq.(4).
Now the eigenenergies are determined mainly by the potential. So it is expected that when $|\alpha|\geq1$, the eigenstates with large eigenenergies are localized states. Another intensively investigated example of  unbounded operator in one dimension  is the Maryland model  where all the eigenstates are localized  \cite{Grempel1982,SIMON,MARX}.
 Furthermore, due to the ergodicity of the map $\phi\longrightarrow 2\pi\beta i+\phi$, for a given sufficiently large real number $\tilde{E}$, there exist some  $i$, the potential $\frac{2\lambda \cos(2\pi\beta i+\phi)}{1-\alpha \cos(2\pi\beta i+\phi)}$ can be very near the real number $\tilde{E}$.
Consequently, there also exists an eigenengy $E_n$ which would be also very near the real number $\tilde{E}$. To be more precise, for an arbitrarily small real number $\delta>0$, there exists
 a real number $M_\delta>0$ ($M_\delta$ usually depends on $\delta$),
when $|\tilde{E}|>M_\delta$, and  there exists an eigenenergy $E_n$, such that the relation
\begin{align}\label{10}
|E_n-\tilde{E}|<\delta
\end{align}
holds.  Roughly speaking, there always exists an eigenenergy in a small neighborhood of a large real number.
 In  this sense, we would say the set of eigenenergies is asymptotically dense in real number set $R$.

\section{The transfer matrix and  the Lyapunov exponent  }
 The  localized properties of eigenstates  can be characterized by the Lyapunov exponent. In this section,  we present the transfer matrix method and its relation to the Lyapunov exponent.

First of all,  we assume the system is a half-infinite lattice system with left-hand end sites $i=0$ and $i=1$.
The Lyapunov exponent can be calculated with the transfer matrix method \cite{Sorets1991,Davids1995}.
 For example, using Eq.(2), starting from  $\psi(0)$ and $\psi(1)$ of left-hand end sites, the wave function can be obtained with relation
\begin{align}\label{V}
\Psi(i)=T(i)T(i-1)...T(2)T(1)\Psi(0)
\end{align}
where matrix
\begin{align}\label{V}
T(n)\equiv\left[\begin{array}{ccc}
\frac{E}{t}-\frac{2\lambda}{t}\frac{ \cos(2\pi\beta n+\phi)}{1-\alpha \cos(2\pi\beta n+\phi)} &-1  \\
1&0\\
  \end{array}\right].
\end{align}
and
\begin{align}\label{V}
\Psi(n)\equiv\left[\begin{array}{ccc}
\psi(n+1)  \\
\psi(n)\\
  \end{array}\right].
\end{align}
If one views Eq.(11) as an evolution equation of dynamical system, $\psi(0)$ and $\psi(1)$ would play the roles of the initial conditions.


For a given real number $E$, with the increasing of $n$, one can assume that the wave function would grow roughly according to an exponential law \cite{Ishii,Furstenberg}, i.e.,
\begin{align}\label{V}
\psi(n)\sim e^{\gamma(E) n}, &\ as \ n\rightarrow \infty,
\end{align}
where $\gamma(E)\geq0$ is Lyapunov exponent which measures the average growth rate of wave function. If the parameter $E$ is not an eigen-energy of $H$, the Lyapunov exponent would be positive, $\gamma(E)>0$ \cite{Jonhnson1986}.
When the parameter $E$ is an eigen-energy of $H$, the Lyapunov exponent can be zero or positive.
For extended states (and critical states), the Lyapunov exponent $\gamma(E)\equiv0$. While for localized states, the Lyapunov exponent $\gamma(E)>0$.

Consequently, the Lyapunov exponent can be written as
\begin{align}\label{V}
&\gamma(E)=\lim_{L \rightarrow \infty }\frac{\ln(|\Psi(L)|/|\Psi(0)|)}{L}\notag\\
&=\lim_{L\rightarrow \infty}\frac{\ln(|T(L)T(L-1)...T(2)T(1)\Psi(0)|/|\Psi(0)|)}{L}
\end{align}
where $L$ is a positive integer and
\begin{align}
|\Psi(n)|=\sqrt{|\psi(n+1)|^2+|\psi(n)|^2}.
\end{align}

The transfer matrix (12) can be further
written as a product of two parts, i.e., $T(n)= A_nB_n$, where
   \begin{align}\label{H0}
&A_n=\frac{1}{1-\alpha \cos(2\pi\beta n+\phi)},\notag\\
&B_n=\left[\begin{array}{ccc}
B_{11} &B_{12} \\
B_{21}&0\\
  \end{array}\right],
\end{align}
with $B_{11}=\frac{E}{t}[1-\alpha \cos(2\pi\beta n+\phi)]-2\lambda \cos(2\pi\beta n+\phi)/t$, and $B_{21}=-B_{12}=1-\alpha \cos(2\pi\beta n+\phi)$.
Now  the Lyapunov exoponent is
\begin{align}\label{H0}
\gamma(E)=\gamma_A(E)+\gamma_B(E),
\end{align}
where
\begin{align}\label{H0}
&\gamma_A(E)=\lim_{L\rightarrow \infty}\frac{\ln(|A(L)A(L-1)...A(2)A(1)|)}{L}.
\end{align}
and \begin{align}\label{V}
&\gamma_B(E)=\lim_{L\rightarrow \infty}\frac{\ln(|B(L)B(L-1)...B(2)B(1)\Psi(0)|/|\Psi(0)|)}{L}.
\end{align}

When $|\alpha|<1$, the quasi-periodic potential is bounded and non-singular.
The Avila's global theory would apply for such a case \cite{Avila2015}.
If parameter $E$ is an eigenvalue of  Hamiltonian $H$, the Lyapunov exponent can be obtained with Avila's  theory \cite{Liu2021}.
When $|\alpha|\geq1$, the Hamiltonian operator is unbounded due to the divergence of potential. Some conclusions of Avila's theory would not be valid for the unbounded case (see next section). Nevertheless,  we would adopt a similar procedure to get the Lyapunov exponent and the Avila's acceleration (see next section). Their correctness would be verified by numerical calculations.

\begin{figure}
\begin{center}
\includegraphics[width=1.0\columnwidth]{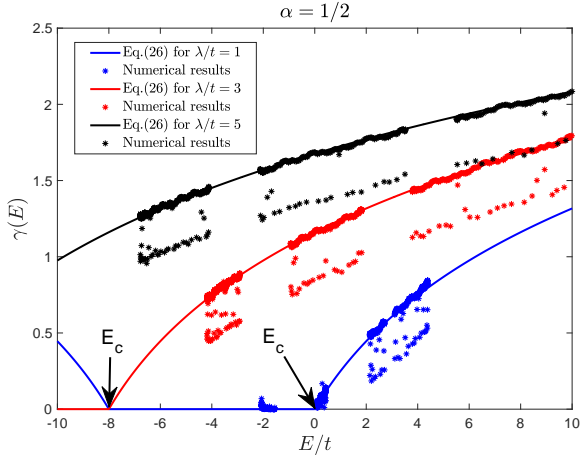}
\end{center}
\caption{ Lyapunov exponents for bounded  case ($\alpha=1/2$) and potential strength $\lambda/t=1,3,5$.  The discrete points are the numerical results for eigenenergies. The solid lines are given by Eq.(26). The mobility edges for $\lambda/t=1$ are indicated by black arrows.  Near mobility edges of the localized-extended transition (e.g., $E_c=0$ and $-8t$ for $\lambda/t=1$), the Lyapunov exponent $\gamma(E)\propto |E-E_c|$ approaches zero (as $E\rightarrow E_c$). The critical index of the localized length $\nu=1$. }
\label{schematic}
\end{figure}

\begin{figure}
\begin{center}
\includegraphics[width=1.0\columnwidth]{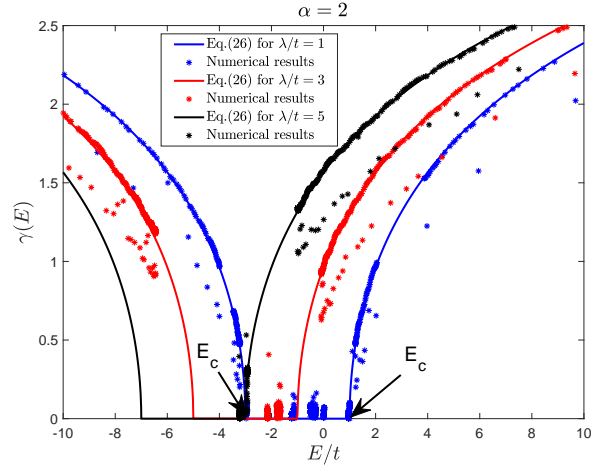}
\end{center}
\caption{ Lyapunov exponents for  unbounded case ($\alpha=2$) and potential strength $\lambda/t=1,3,5$. The discrete points are the numerical results for eigenenergies. The solid lines are given by Eq.(26). The mobility edges for $\lambda/t=1$ are indicated by black arrows.  Near the localized-critical  transition (e.g., $E_c=t$ and $-3t$ for $\lambda/t=1$), the Lyapunov exponent $\gamma(E)\propto |E-E_c|^{1/2}$ (as $E\rightarrow E_c$), and the critical index of the localized length $\nu=1/2$. }
\label{schematic}
\end{figure}

Following Refs.\cite{Liu2021,YONGJIAN1,YONGJIAN2}, first of all, we  complexify the phase $\phi\rightarrow \phi+i \epsilon$ with $\epsilon >0$ , e.g., $B_{11}=\frac{E}{t}[1-\alpha \cos(2\pi\beta n+\phi+i \epsilon)]-2\lambda \cos(2\pi\beta n+\phi+i \epsilon)/t$, and $B_{21}=-B_{12}=1-\alpha \cos(2\pi\beta n+\phi+i \epsilon)$.
 In addition, due to the ergodicity of the map $\phi\longrightarrow 2\pi\beta n+\phi$, we can write  $\gamma_A(E)$ as integral over phase $\phi$ \cite{Longhi2019}, consequently
\begin{align}\label{H0}
&\gamma_A(E,\epsilon)=\frac{1}{2\pi}\int_{0}^{2\pi} \ln(\frac{1}{|1-\alpha \cos(\phi+i\epsilon)|})d\phi\notag\\
&=\left\{\begin{array}{ccc}
-\ln(\frac{1+\sqrt{1-\alpha^2}}{2}),  & for \ |\alpha|<1 \ \& \ \ \epsilon<\ln(\frac{1+\sqrt{1-\alpha^2}}{|\alpha|})\\
-\epsilon-\ln(\frac{|\alpha|}{2}),&for  \ |\alpha|\geq1.
  \end{array}\right.
\end{align}

Next we take $\epsilon\rightarrow\infty$
 \begin{align}\label{H0}
B_n=\frac{e^{-i(2\pi\beta n+\phi)+\epsilon}}{2}\left[\begin{array}{ccc}
\frac{-(\alpha E+2\lambda)}{t} &\alpha \\
-\alpha&0\\
  \end{array}\right]+O(1).
\end{align}
Then for large $\epsilon$, i.e.,  $\epsilon\gg1$, $\gamma_B(E,\epsilon)$ is determined by the largest eigenvalue (in absolute value) of $B_n$, i.e.,
\begin{align}\label{H0}
&\gamma_B(E,\epsilon)=
\left\{\begin{array}{cc}
\epsilon+\ln(\frac{|P|+\sqrt{P^2-4\alpha^2}}{4}),  & for \ P^2>4\alpha^2\\
\epsilon+\ln(\frac{|\alpha|}{2}),&for \ P^2<4\alpha^2,
  \end{array}\right.
\end{align}
where
\begin{align}\label{H0}
P=\frac{\alpha E+2\lambda}{t}.
\end{align}

When $\epsilon$ is very small, using the facts that $\gamma(E,\epsilon)\geq0$  and $\gamma(E,\epsilon)$ is  a convex and piecewise linear function of $\epsilon$ \cite{Avila2015,YONGJIAN2},  one can get
\begin{align}\label{H0}
&\gamma(E,\epsilon)=Max\{0,\gamma_A(E,\epsilon)+\gamma_B(E,\epsilon)\},\notag\\
&=\left\{\begin{array}{cccc}
Max\{0,\epsilon+\ln(\frac{|P|+\sqrt{P^2-4\alpha^2}}{2|1+\sqrt{1-\alpha^2}|})\},   \ |\alpha|<1 \ \& \ P^2>4\alpha^2\\
0, \ \ \ \ \ \ \ \ \ \ \ \ \ \ \ \ \ \ \ \ \ \ \ \ \ \ \ \ \ \ \ \ \ |\alpha|<1 \ \& \ P^2<4\alpha^2\\
\ln(\frac{|P|+\sqrt{P^2-4\alpha^2}}{2|\alpha|}),\ \ \ \ \ \ \ \ \ \ \ \ |\alpha|\geq1 \ \& \ P^2>4\alpha^2\\
0.\ \ \ \ \ \ \ \ \ \ \ \ \ \ \ \ \ \ \ \ \ \ \ \ \ \ \ \ \ \ \ \ \ \ |\alpha|\geq1 \ \& \ P^2<4\alpha^2.
  \end{array}\right.
\end{align}
Furthermore, when $\epsilon=0$, the Lyapunov exponent $\gamma(E)\equiv\gamma(E,\epsilon=0) $ is
\begin{align}\label{Coul}
&\gamma(E)=
\left\{\begin{array}{cccc}
Max\{0,\ln(\frac{|P|+\sqrt{P^2-4\alpha^2}}{2|1+\sqrt{1-\alpha^2}|})\},   \ |\alpha|<1 \ \& \ P^2>4\alpha^2\\
0, \ \ \ \ \ \ \ \ \ \ \ \ \ \ \ \ \ \ \ \ \ \ \ \ \ \ \ \ \ \ \ \ \ |\alpha|<1 \ \& \ P^2<4\alpha^2\\
\ln(\frac{|P|+\sqrt{P^2-4\alpha^2}}{2|\alpha|}),\ \ \ \ \ \ \ \ \ \ \ \ |\alpha|\geq1 \ \& \ P^2>4\alpha^2\\
0.\ \ \ \ \ \ \ \ \ \ \ \ \ \ \ \ \ \ \ \ \ \ \ \ \ \ \ \ \ \ \ \ \ \ |\alpha|\geq1 \ \& \ P^2<4\alpha^2.
  \end{array}\right.
\end{align}
The above generalized formula Eq.(26) for both $|\alpha|<1$ and $|\alpha|\geq1$  has been  verified by our numerical results (see Figs.1 and 2).

In our numerical calculations, in order to get the correct Lyapunov exponents, on the one hand, the integer $L$ should be sufficiently large.
On the other hand, $L$ should be also much smaller than the system size $N$, i.e., $1\ll L\ll N$.
 To be specific, taking $\alpha=1/2,2$, $\lambda/t=1,3,5$, system size $N=1000$, we get the $N=1000$ eigenenergies and eigenstates.
  Then, we calculate the Lyapunov exponents numerically for all the eigenenergies [see the several sets of discrete points in  Figs.1 and 2].
 In our numerical calculation, we take $L=200$,  phase $\phi=0$, $\psi(0)=0$ and $\psi(1)=1$ in Eq.(15).
  The solid lines of Figs.1 and 2 are given by Eq.(26) with same parameters. It is shown that most of all discrete points fall onto the solid lines.

However, we also note that there are some discrete points of localized states which are not on the solid lines. This is because these localized wave functions  are too near the left-hand boundary of system.

\section{the mobility edges and  critical region }
In this section, based on the Lyapunov exponent formula Eq.(26), we determine the mobility edges and the critical region.
\subsection{$|\alpha|<1$}
When $|\alpha|<1$, there exist localized-extended transitions \cite{Ganeshan2015} (see Fig.3). Based on the Eq.(26), the mobility edges $E_c$ which separate the localized from the extended states, are determined by \cite{Liu2021}
\begin{align}\label{H0}
\gamma(E_c)=\ln(\frac{|P|+\sqrt{P^2-4\alpha^2}}{2|1+\sqrt{1-\alpha^2}|})=0
\end{align}
then
\begin{align}\label{H0}
|P|=2\rightarrow |\frac{\alpha E_c+2\lambda}{t}|=2,
\end{align}
which is consistent with Ganeshan et al's result \cite{Ganeshan2015} for $|\alpha|<1$.
 Furthermore, when $\alpha=0$, the transition point is given by
\begin{align}\label{H0}
|\lambda|=|t|,
\end{align}
which is reduced into the well-known Aubry-Andr\'{e}'s self-dual result \cite{Aubry1980}.

By expanding the Lyapunov exponent near the mobility edges $E_c$, we get
\begin{align}\label{V}
\gamma(E)\propto |E-E_c|\rightarrow0, \ as \ E \rightarrow E_c.
\end{align}
Then the localized length is
\begin{align}\label{V}
\xi(E)\equiv1/\gamma(E)\propto |E-E_c|^{-1}\rightarrow\infty,\ as \ E \rightarrow E_c.
\end{align}
Its critical index is $1$ for the bounded  case of $|\alpha|<1$ [see the finite slopes of solid lines near $E_c$ in Fig.1], which is also consistent with the numerical findings \cite{Xiaopeng,Deng}.
When  $E$ is an eigenenergy and $E=E_c$, the state of $E$ is a critical state. Because the energy $E=E_c$ is an isolated point of real number set $R$, the critical states  at $E=E_c$ are usually unstable under perturbations \cite{Avila2015}.

\begin{figure}
\begin{center}
\includegraphics[width=1.0\columnwidth]{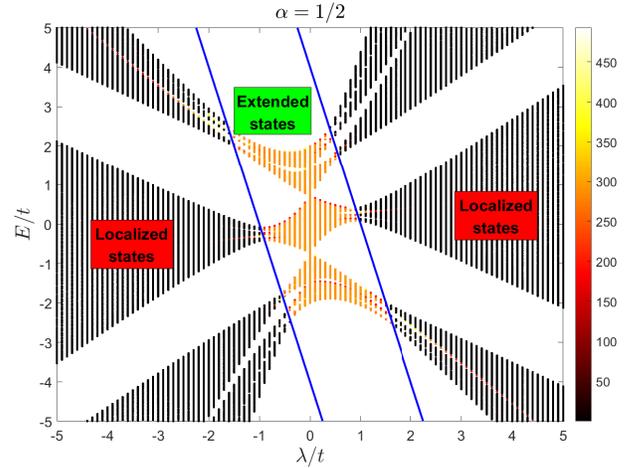}
\end{center}
\caption{ Phase diagram in $(\lambda, E)$ plane for bounded case ($\alpha=1/2$). When $\alpha=1/2$, there exists localized-extended transitions. The blue solid lines are the phase boundaries (mobility edges $E_c$), which  are given by Eq.(28).
 Standard deviations are represented with different colors. }
\label{schematic}
\end{figure}

\begin{figure}
\begin{center}
\includegraphics[width=1.0\columnwidth]{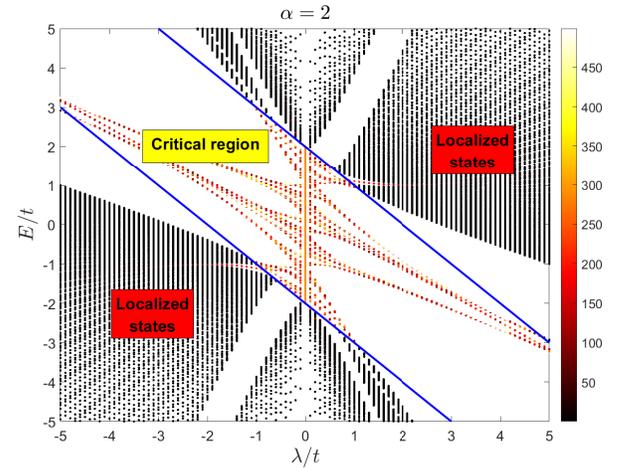}
\end{center}
\caption{ Phase diagram in $(\lambda, E)$ plane for unbounded case ($\alpha=2$). When $\alpha=2$, there exist localized-critical transitions. The blue solid lines are the phase boundaries (mobility edges $E_c$), which  are given by Eq.(33).
 Standard deviations are represented with different colors. Within the critical region, there are large fluctuations in standard deviations.}
\label{schematic}
\end{figure}

\begin{figure}
\begin{center}
\includegraphics[width=1.1\columnwidth]{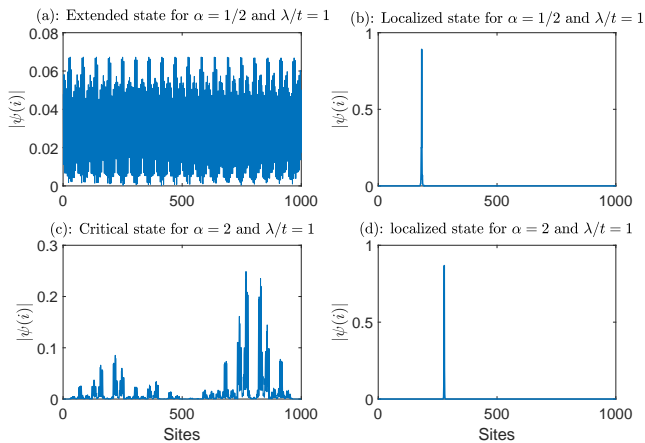}
\end{center}
\caption{ Several typical wave functions for extended, localized, and critical states.}
\label{schematic}
\end{figure}

\subsection{$|\alpha|\geq1$}
When $|\alpha|\geq1$, there are localized-critical transitions (see Fig.4). The mobility edges $E_c$ which separate the localized from the critical states, by the Eq.(26),  are determined by
\begin{align}\label{H0}
P^2=4\alpha^2
\end{align}
then
\begin{align}\label{H0}
|P|=2|\alpha|\rightarrow |\frac{\alpha E_c+2\lambda}{t}|=2|\alpha|.
\end{align}
The critical region (see Fig.4) is given by
\begin{align}\label{H0}
|P|<2|\alpha|\rightarrow |\frac{\alpha E+2\lambda}{t}|<2|\alpha|.
\end{align}
Near the mobility edges $E_c$, we find that the Lyapunov exponent behaves as
\begin{align}\label{V}
\gamma(E)\propto |E-E_c|^{1/2}\rightarrow 0,\ as \ E \rightarrow E_c.
\end{align}
Then the localized length is
\begin{align}\label{V}
\xi(E)\equiv1/\gamma(E)\propto |E-E_c|^{-1/2}\rightarrow\infty, \ as \ E \rightarrow E_c.
\end{align}
Its critical index is $1/2$ [see the infinitely large slopes of solid lines near $E_c$ in Fig.2].

Several typical wave functions for localized, critical  and extended states are reported in Fig.5.
We can see that the wave function of extended state extends all over the whole lattices, while localized state only occupies finite lattice sites.
The critical state consists of several disconnected patches which interpolates between the localized and extended states.
In comparison with the extended states, there exist some unoccupied regions in critical state wave function.

In order to further distinguish the localized states from the extended states (and critical states), we also numerically calculate standard deviation of coordinates of eigenstates \cite{Boers2007}
\begin{align}\label{37}
&\sigma=\sqrt{\sum_{i}(i-\bar{i})^2|\psi(i)|^2},
\end{align}
where the average value of coordinate $\bar{i}$ is
\begin{align}\label{Coul}
\bar{i}=\sum_{i}i|\psi(i)|^2.
\end{align}
The standard deviation $\sigma$ describes the spatial extension of wave function in the lattices. If one views $E$ as a parameter, a ``phase diagram" in $(\lambda, E)$ plane can be obtained.
The phase diagram is reported in Figs.3 and 4.  In Figs.3 and 4, the standard deviations of coordinates are represented with different colors.
From Figs.3 and 4, we can see that when the states are localized, standard deviations of coordinates are very small. For extended states, the standard deviations are very large.
The standard deviations of the critical states are in between of them (also see Figs.6 and 7 and Table \textbf{II}).

\begin{figure}
\begin{center}
\includegraphics[width=1.0\columnwidth]{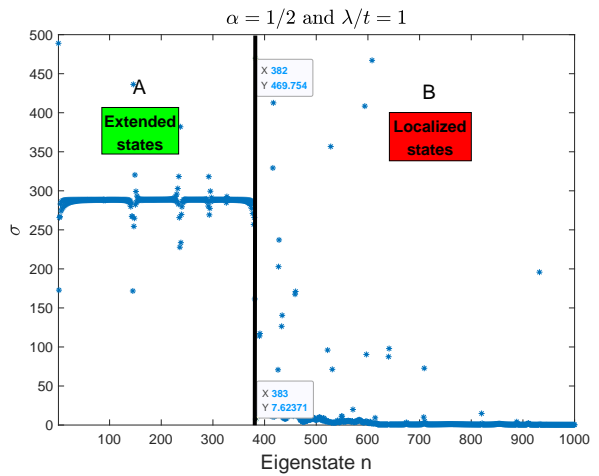}
\end{center}
\caption{ Standard deviations of localized states and extended states for parameter $\alpha=1/2$ and $\lambda/t=1$. The eigenenergy $E_n$ increases gradually as eigenstate index $n$ runs from $1$ to $1000$.}
\label{schematic}
\end{figure}

\begin{figure}
\begin{center}
\includegraphics[width=1.0\columnwidth]{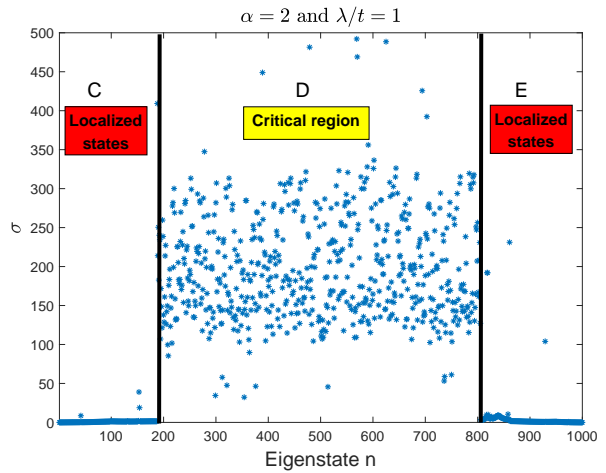}
\end{center}
\caption{  Standard deviations of localized states and extended states for parameter $\alpha=2$ and $\lambda/t=1$.  The eigenenergy $E_n$ increases gradually as eigenstate index $n$ runs from $1$ to $1000$.}
\label{schematic}
\end{figure}

For a given potential strength $\lambda/t=1$, we report the standard deviations of eigenstates in Figs.6 and 7. It is shown that in comparison with extended states and localized states, the critical states have much larger fluctuations of standard deviations.
In order to see their differences, we calculate  the fluctuation $f_\Omega$ for a given set of eigenstates $\Omega$
\begin{align}\label{Coul}
f_\Omega\equiv\sqrt{\sum_{k\in \Omega} (\sigma_k-\bar{\sigma}_\Omega)^2/N_\Omega}
\end{align}
where $N_\Omega$ is total eigenstate number in set $\Omega$ and the average  value of standard deviations
\begin{align}\label{Coul}
\bar{\sigma}_\Omega=\frac{1}{N_\Omega}\sum_{k\in \Omega} \sigma_k.
\end{align}
\
\begin{table}
\begin{center}
\begin{tabular}{|c|c|c|c|}
\hline
\hline
& Extended states &Localized states &Critical states \tabularnewline
\hline
 &$\alpha=1/2$ \& $\lambda/t=1$& $\alpha=2$ \& $\lambda/t=1$&  $\alpha=2$ \& $\lambda/t=1$    \tabularnewline
 \hline
 $\bar{\sigma}_\Omega$ &287.94& 1.19 &  199.77  \tabularnewline
 \hline
 $f_\Omega$& 20.78&3.16 & 68.86   \tabularnewline
 \hline
\end{tabular}
\end{center}
\caption{the average values $\bar{\sigma}$ and its fluctuations for extended, localized and critical states. The sets of extended states, localized states and critical states correspond regions A, C and D of Figs.6 and 7, respectively.
}
\end{table}
When $\alpha=1/2$ and $\lambda/t=1$,  we take the set of extended states $\Omega_E$ where the state number runs from 1 to 382, i.e, region \textbf{A} of Fig.6.
When $\alpha=2$ and $\lambda/t=1$,  we take the set of localized states $\Omega_L$ where the state number runs from 1 to 187, i.e, region \textbf{C} of Fig.7.
For critical states, we take region \textbf{D} of Fig.7 as set of eigenstates $\Omega_{Cr}$ .
The results are reported in Table \textbf{II}. It is shown that the fluctuation of critical states is  much larger than that of  the localized and extended states.

\begin{figure}
\begin{center}
\includegraphics[width=1.1\columnwidth]{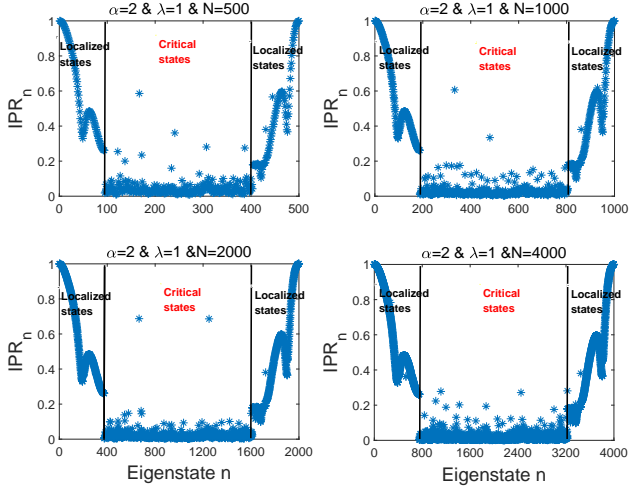}
\end{center}
\caption{ The inverse
participation ratio $IPR_n$ of all the eigenstates for system size $N=500,1000, 2000$ and $N=4000$.}
\label{schematic}
\end{figure}

In order to investigate the properties of  the wave functions of critical states, we also numerically calculate the inverse participation ratio $IPR_n$ of  all eigenstates for different system sizes $N=500,1000,2000$ and $N=4000$ \cite{Xiaopeng,Deng}, i.e.,
\begin{align}
&IPR_n=\sum_{i}|\psi_n(i)|^4.
\end{align}
where $\psi_n(i)$ is the normalized wave function for $n-th$ eigenstate.
The results are reported in Fig.8.
We find that the IPRs of localized states are basically same for different system sizes $N$,  while the IPRs of critical states have much larger fluctuations.

\begin{figure}
\begin{center}
\includegraphics[width=1.1\columnwidth]{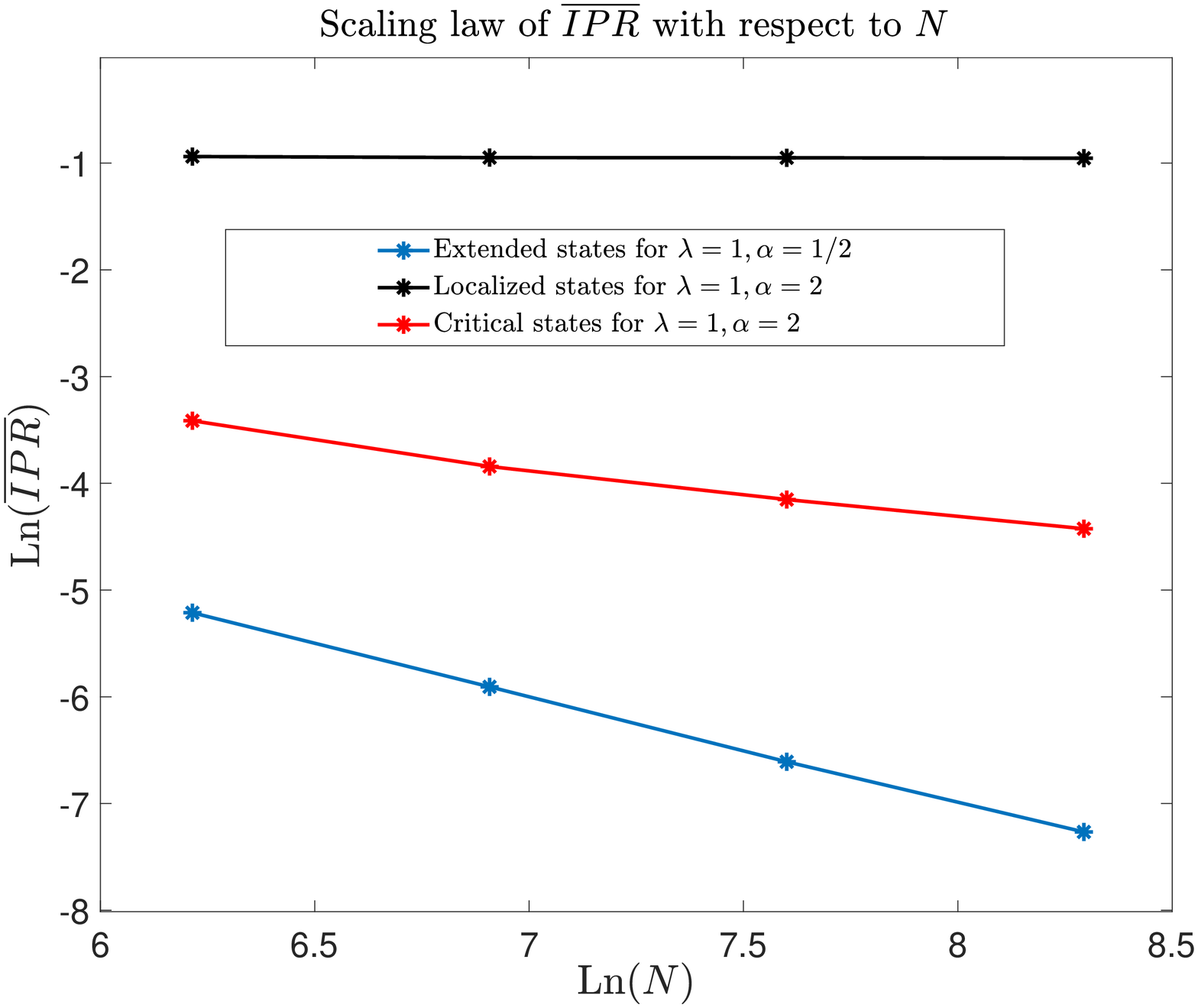}
\end{center}
\caption{The scaling law of $\overline{IPR}$.
We calculate the average participation ratio $\overline{IPR}$ for the eigenstates in some typical energy intervals. The energy intervals for localized states, critical states and extended states are $-3.5<E/t<-3.2$, $-0.5<E/t<-0.3$, and $-2.1<E/t<-1.5$, respectively.
The system sizes for the discrete points are $N=500,1000, 2000$ and $N=4000$, respectively. It is found that  for localized states, the scaling exponent $x\simeq0$.  While for extended states, the scaling exponent $x\simeq1$.  For critical states, the scaling exponent  $x\simeq 0.5$.}
\label{schematic}
\end{figure}

On the whole,   the IPR of critical states decreases with the increasing of system size $N$.
 The decreasing law may be captured by a power law function
\begin{align}\label{H0}
\overline{IPR}\propto 1/N^x,
\end{align}
where $\overline{IPR}$ is an average value of the  $IPR$ within a typical energy interval, and $x$ is scaling exponent.
 It is believed  that, for localized states, the scaling exponent $x=0$.  While for extended states (like plane wave states),  the scaling exponent $x=1$. For critical states, the exponent should be $0<x<1$ (see Fig.9).
For different system sizes $N$, due to randomness of $IPR$ of critical states (see Fig.8),  it is difficult to get a definite scaling exponent $x$.
Here we find that for the critical states in the energy interval $-0.5<E/t<-0.3$, the scaling exponent  satisfies $0.39<x<0.62$, its average value $\bar{x}\simeq0.5$ (see Fig.9).

\subsection{Avila's acceleration}
In addition, for the bounded quasi-periodic potentials, Avila also defined the acceleration  $\omega(E)$ by \cite{Avila2015}
\begin{align}\label{H0}
\omega(E)=lim_{\epsilon\rightarrow 0}\frac{\gamma(E,\epsilon)-\gamma(E,0)}{\epsilon}.
\end{align}
Furthermore it is proved that acceleration $\omega(E)\geq0$ and is quantized (an integer) for a bounded operator $H$. For critical states of $E=E_c$, $\gamma(E)=0$ and $\omega(E)\neq0$.
Similarly, using Eqs.(25), when real number $E$ is an eigenvalue of $H$,  we extend it into the case of $|\alpha|\geq1$ by
\begin{align}\label{Coul}
&\omega(E)=
\left\{\begin{array}{cccc}
1, \ for  \ |\alpha|<1 \ \& \ energy \ of \ localized \ state \\
0, \ for  \ |\alpha|<1 \ \& \ energy \ of \ extended \ state \\
1, \ \ \ for  \ |\alpha|<1 \ \& \ energy \ of \ critical \ state \\
0,\ \ \ \ \ \ \ \ for  \ |\alpha|\geq 1 \ \& \ if \ E \ is \ eigenenergy
  \end{array}\right.
\end{align}
We note that that the second part of  Eq.(21) is a  linear function of $\epsilon$, while the first part does not depend on $\epsilon$.
This is why  accelerations $\omega(E)$ for $|\alpha|<1$ and $|\alpha|\geq1$ are different.

\begin{figure}
\begin{center}
\includegraphics[width=1.1\columnwidth]{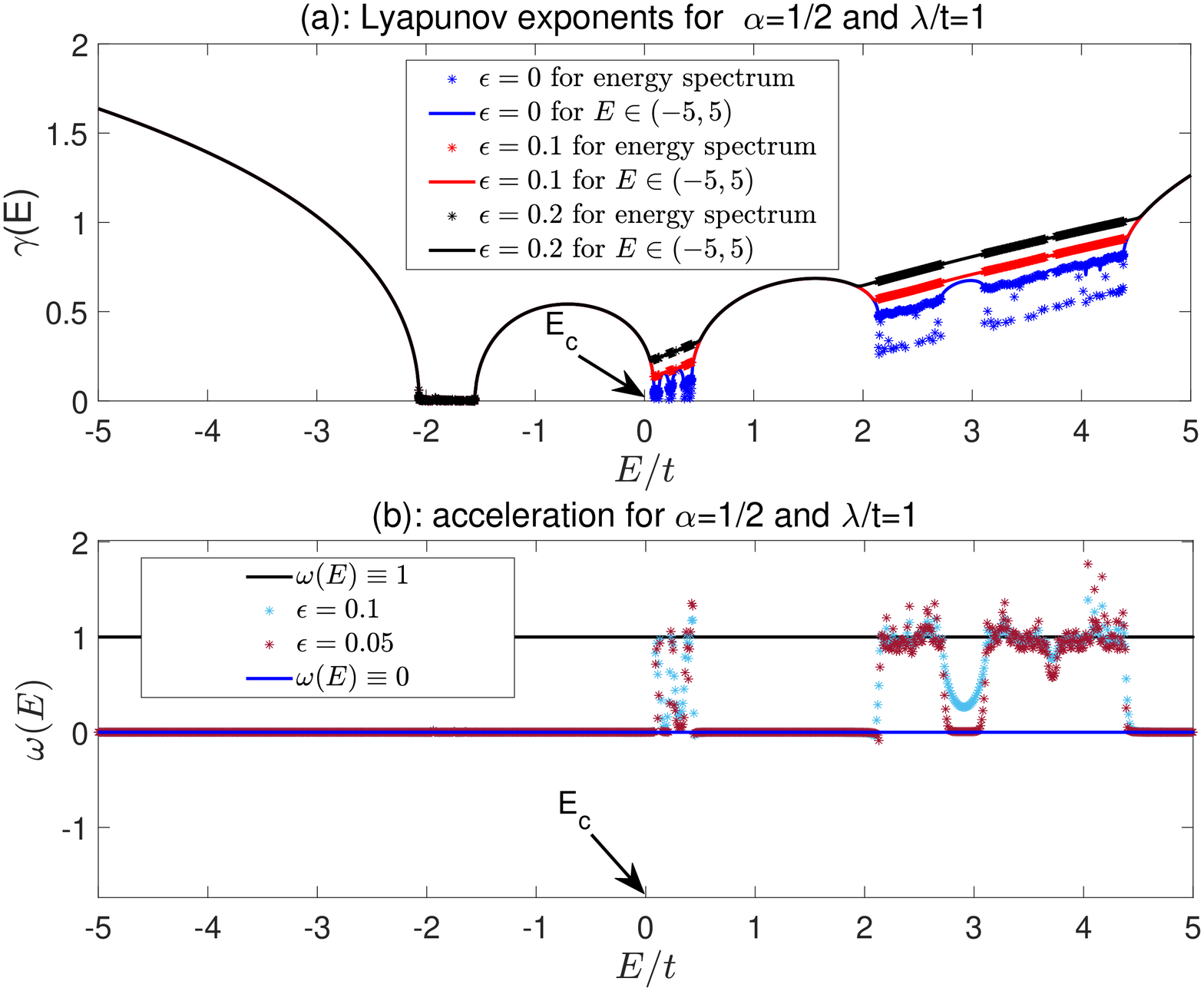}
\end{center}
\caption{ Lyapunov exponents and Avila's accelerations for  localized states and extended states. (a): Lyapunov exponents for $\alpha=1/2$ and $\lambda/t=1$.   (b): Avila's accelerations  for $\alpha=1/2$. The mobility edges $E_c=0$ are indicated by black arrows in figure. }
\label{schematic}
\end{figure}

\begin{figure}
\begin{center}
\includegraphics[width=1.1\columnwidth]{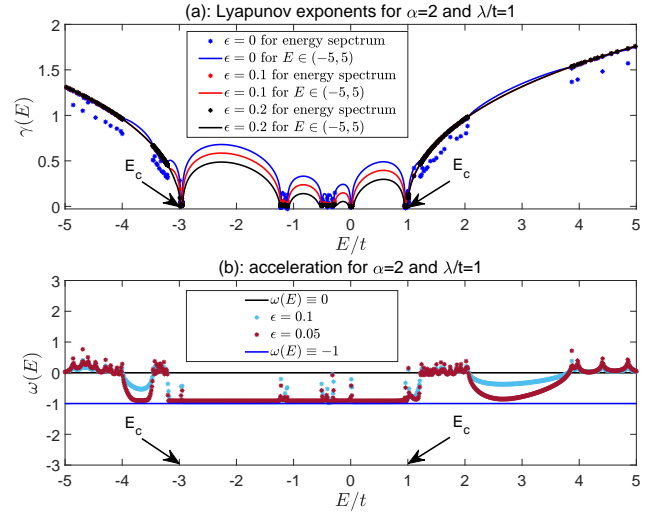}
\end{center}
\caption{ Lyapunov exponents and Avila's accelerations for  localized states and extended states. (a): Lyapunov exponents for $\alpha=2$ and $\lambda/t=1$.(b): Avila's accelerations  for $\alpha=2$.  The mobility edges $E_c=-3t$  and $t$ are indicated by black arrows in figure.  }
\label{schematic}
\end{figure}

When $|\alpha|<1$, by Avila's global theory \cite{Avila2015}, for an analytical (bounded) quasi-periodic potential, if the real number $E$ is not an eigen-value of Hamiltonian $H$, the Lyapunov exponent is always positive, i.e., $\gamma(E)>0$ and the acceleration is always zero, i.e., $\omega(E)\equiv0$. Further combining Eq.(26) and Eq.(44),
then one can classify systems with different real parameter $E$ (different phases) by Lyapunov exponent and the quantized acceleration, i.e.,
\begin{align}\label{Coul}
 &(a): \gamma(E)>0 \ \ \& \ \omega(E)=0,  \ if \ E \ is \ not\ an\ eigenvalue \notag\\
 &(b): \gamma(E)>0 \ \ \& \ \omega(E)=1,  \ for \ localized \ state\notag\\
 &(c): \gamma(E)=0 \ \ \& \ \omega(E)=0,  \ for \  extended \ state\notag\\
 &(d): \gamma(E)=0 \ \ \& \ \omega(E)=1,    \ for \ critical \ state.
\end{align}
The above results are verified by our numerical calculations (see Fig.10).
To be specific, taking $\alpha=1/2$, $\lambda/t=1$ and $\epsilon=0,0.1,0.2$, we calculate the Lyapunov exponents  with Eq.(15) (taking the complexified phase $\phi\rightarrow \phi+i\epsilon=i\epsilon$) for interval $-5\leq E\leq5$ [see the three solid lines in panel (a) of Fig.10]. In our calculation, we take $L=200$, $\psi(0)=0$ and $\psi(1)=1$ in Eq.(15).
At the same time, we calculate the Lyapunov exponents for all the eigenenergies with same parameters [see the three sets of discrete points in panel (a) of Fig.10].
We can find that if $E$  is not an eigenenergy, its Lyapunov exponents are the same for all three different  $\epsilon=0,0.1,0.2$.
When $E$ is an eigenenergy of extended state [$\gamma(E)=0$], the Lyapunov exponents are also the same for all three different  $\epsilon=0,0.1,0.2$.
While when $E$ is an eigenenergy of localized state [$\gamma(E)>0$], the Lyapunov exponents are different for  three different  $\epsilon=0,0.1,0.2$. Their differences are linearly proportional to $\Delta\epsilon=0.1$ in Fig. 10.

By taking $\epsilon=0.1$ and $\epsilon=0.05$, we also  approximately calculate Avila's  acceleration $\omega(E)$ by
\begin{align}\label{H0}
\omega(E)\simeq\frac{\gamma(E,\epsilon)-\gamma(E,0)}{\epsilon},
\end{align}
[see panel (b) of Fig.10]. It shows that when $E$ is an eigenenergy of localized state [$\gamma(E)>0$],
Avila's  acceleration is 1. Otherwise,
Avila's  acceleration is 0.

Next we also carry a similar calculations for the case of $\alpha=2$  $(|\alpha|\geq1)$ in panels (a) and (b) of Fig.11. It is found that when $E$ is not an eigenenergy, Avila's  acceleration is $-1$. For other cases, Avila's  acceleration is always $0$.
Consequently, for $|\alpha|\geq1$, the systems with different real number $E$ can be classified by
\begin{align}\label{Coul}
 &(a): \gamma(E)>0 \ \ \& \ \omega(E)=-1,  \ if \ E \ is \ not\ eigenvalue \notag\\
 &(b): \gamma(E)>0 \ \ \& \ \omega(E)= 0,  \ for  \ localized \ state\notag\\
 &(c): \gamma(E)=0 \ \ \& \ \omega(E)= 0,   \ for \  critical \ state\notag\\
 &(d): \gamma(E)=0 \ \ \& \ \omega(E)\neq 0,  \ such  \ E \ does\ not\ exist.
\end{align}
It is  noted that Avila's acceleration is also quantized for the unbounded quasi-periodic potential in the GAA model.
 From Eqs.(45) and (47), we see Avila's acceleration can be used to distinguish the case $(a)$ from case $(b)$ of real number $E$.

\section{summary}
In conclusion, we extend the investigations of GAA model into a regime of parameter $|\alpha|\geq1$.
 It is found that there exist mobility edges which separate the localized states from critical states.
 Within the critical region, the spatial extensions of eigenstates have large fluctuations.

 The Lyapunov exponents and  mobility edges are  exactly obtained with Avila's theory for both $|\alpha|<1$ and $|\alpha|\geq1$ cases.
 Furthermore, it is found that the critical index of localized length $\nu=1$ for $|\alpha|<1$, while for $|\alpha|\geq1$, $\nu=1/2$.
The two different critical indices can be used to distinguish the localized-extended transitions from  localized-critical transitions.
The numerical results show that the scaling exponent of inverse participation ratio (IPR) of critical states $x\simeq0.5$.
In addition,  it is shown that the Lyapunov exponent and Avila's acceleration can be used to classify the systems with different $E$ for  both $|\alpha|<1$ and $|\alpha|\geq1$.

In some sense, we extend  Avila's theory to unbounded quasi-periodic potentials in the GAA model. For example, we find that if $E$ is not an eigenenergy, Avila's acceleration $\omega(E)=-1<0$
 which is different from  Avila's prediction [$\omega(E)\geq0$] for bounded quasi-periodic potentials. In addition,  when $E$ is an eigenenergy of a localized state, it is found that $\omega(E)=1$ for bounded quasi-periodic potentials which is consistent with Avila's theory [$\omega(E)$ is a positive integer]. While for unbounded quasi-periodic potential,  we find that $\omega(E)=0$ for localized states, which is also different from bounded potential case.
A much more exact theory for the unbounded quasi-periodic potential needs further investigations. We anticipate the work will spark further interests in the exact localization theory for the unbounded quasi-periodic potentials.


\section*{Acknowledgements}
This work was supported by the NSFC under Grants Nos.
11874127, 11774336.


\end{document}